\documentstyle[aps,preprint,epsfig,floats]{revtex}   % REVTeX v.3.0. For double spacing

\begin{document}

\def\sref#1{\protect\ref{#1}}
\def\fref#1{\protect\ref{#1}}
\def\eref#1{(\protect\ref{#1})}

\newcommand{\ground}{{\mbox{\scriptsize{ground}}}}

\def\capt#1{\refstepcounter{figure}\bigskip\hbox to \textwidth{%
       \hfil\vbox{\hsize=5in\renewcommand{\baselinestretch}{1}\small
       {\sc Figure \thefigure}\quad#1}\hfil}\bigskip}

%\draft
%\tolerance = 10000

% Fixing abstract in twocolumn mode 
%\twocolumn[\hsize\textwidth\columnwidth\hsize\csname @twocolumnfalse\endcsname

\title{Glassy dynamics and aging in an exactly solvable spin model}
\author{M. E. J. Newman and Cristopher Moore}
\address{Santa Fe Institute, 1399 Hyde Park Road, Santa Fe, NM 87501}
\maketitle

\begin{abstract}
  We introduce a simple two-dimensional spin model with short-range
  interactions which shows glassy behavior despite a Hamiltonian which is
  completely homogeneous and possesses no randomness.  We solve exactly for
  both the static partition function of the model and the distribution of
  energy barriers, giving us the equilibration time-scales at low
  temperature.  Simulations of instantaneous quenches and of annealing of
  the model are in good agreement with the analytic calculations.  We also
  measure the two-time spin correlation as a function of waiting time, and
  show that the model has aging behavior consistent with the distribution
  of barrier heights.  The model appears to have no sharp glass transition.
  Instead, it falls out of equilibrium at a temperature which decreases
  logarithmically as a function of the cooling time.
\end{abstract}

\vspace{0.5cm}

% Fixing abstract in twocolumn mode 
%]

\section{Introduction}
A great deal of effort has been devoted in the last twenty years or so to
understanding the behavior of spin glasses and other glassy
models\cite{Jackle86,FH91,Parisi94}.  In spin glasses, one introduces
randomness into the Hamiltonian of some otherwise well-behaved system,
creating a hierarchical distribution of energy barriers over state space
which prevents the system from reaching thermal equilibrium on reasonable
time-scales below a certain temperature.  The slow dynamics displayed by
these systems has made their computer simulation very difficult despite the
recent appearance of a number of new and promising
algorithms\cite{MPR97,NB99}, and the presence of randomness in the
Hamiltonian has, except in a few special
cases\cite{SK75,Derrida81,Ritort95}, prevented their exact solution.  As a
result, our understand of their behavior is, even after many years of
effort, still very far from complete.  It is, for example, still an open
question whether, in the limit of infinitely slow cooling, spin glasses
with short-range interactions display a sharp transition from ergodicity to
glassy behavior, or whether the transition is a gradual one\cite{FH91}.

However, it is not necessary to have randomness in the Hamiltonian in order
for a system to be glassy.  Glassiness has its origin in the dynamics by
which the system is updated rather than the energy landscape.  In fact, no
landscape even exists until we specify the dynamics, since the set of
elementary moves by which the system moves from one state to another
defines which states are neighbors.  Given an appropriate choice of
dynamics, any system can be ergodic on short time-scales, regardless of the
energies of particular states.  Conversely, it should be possible to find
systems which display glassy behavior without randomness in the
Hamiltonian.

One such system is the molecular or configurational glass---window glass,
for example---but this is a notoriously difficult system to study
mathematically\cite{Jackle86}.  Recently therefore, a number of authors
have investigated spin models which are non-random but show glassy behavior
either because of competition between different types of
interactions\cite{SHS92} or because of the presence of higher-order
interactions\cite{Rieger92,KRS94}.  For some models with infinite-range
interactions, the statics, though not the dynamics, can be solved exactly
\cite{bernasconi,bouchaud,marinari}.

In this paper we introduce a $p$-spin model in two dimensions which, under
a dynamics which flips single spins, displays the classic features of a
glassy system.  This model possesses the considerable advantage over
previously studied models that both its statics and its low-temperature
dynamics are exactly solvable, even though it has only short-range
interactions.

The structure of the paper is as follows.  In Section~\sref{model} we
define our model.  In Section~\sref{equilibrium} we give an analytic
solution for the partition function and internal energy of the model in
equilibrium.  In Section~\sref{glassy} we solve for the distribution of
energy barriers between the ground state and the lowest-lying excitations
of the model and hence argue that it should display glassy behavior.  We
compare our predictions with extensive Monte Carlo simulations and find
excellent agreement between the two.  In Section~\sref{aging} we study the
aging properties of our model, and in Section~\sref{concs} we give our
conclusions.

\section{The model}
\label{model}
Our model is a $p$-spin model composed of Ising spins $\sigma_1=\pm1$ on a
triangular lattice with short-range interactions and a single-spin-flip
dynamics.  The Hamiltonian is
\begin{equation}
H = \mbox{$\frac{1}{2}$} J\!\! \sum_{i,j,k\ {\rm in}\ \bigtriangledown}\!\!
    \sigma_i \sigma_j \sigma_k.
\label{h1}
\end{equation}
The sum here runs over all sets of three nearest-neighbor spins
$i,j,k$ which lie at the three vertices of one of the
downward-pointing triangles on the lattice.  Except for this
restriction to downward-pointing triangles, our model is the same as
the Baxter-Wu model\cite{BW73}, although its behavior is entirely
different.  It is also similar to a model used by
Barkema~{\it{}et~al.}\cite{BNB94} to study the formation of adatom
islands on $(111)$ surfaces of metals.

For most of our presentation we will find it more convenient to
rewrite this Hamiltonian in the form
\begin{equation}
H = J\!\!\sum_{i,j,k\ {\rm in}\ \bigtriangledown}\! (s_i+s_j+s_k) \bmod 2,
\end{equation}
which is identical to~\eref{h1} except for an additive constant if we map
the Ising spins $\sigma_i$ onto the variables
$s_i=\frac{1}{2}(\sigma_i+1)$, which take the values 0~(down) or 1~(up).

One could also construct a disordered version of the model in which the
three-spin interactions were chosen randomly to have strengths $\pm J$.
However, this disordered version can be mapped onto the homogeneous one
above via a simple gauge transformation, and so the two have identical
behavior.  (This transformation is particularly obvious when viewed in
terms of the defect variables introduced below.)

The dynamics of the model consists of moves which flip single spins.
We have chosen to investigate the behavior of the model under the
standard Metropolis dynamics\cite{MRRTT53} in which moves with energy
cost $\Delta E$ take place with rate $1$ if $\Delta E \le 0$, and with
rate ${\rm e}^{-\beta\,\Delta E}$ if $\Delta E > 0$.  However, except
for differences in the short-time correlations and a possible overall
rescaling of time, we would expect the fundamental properties of the
model to be the same for any other single-spin-flip dynamics which
respects both ergodicity and detailed balance.

\section{Equilibrium solution of the model}
\label{equilibrium}
In the following sections we discuss the glassy behavior of our model.
First, however, we give an exact solution of its equilibrium properties.
An alternative representation of the state of the model is as a triangular
lattice of defects: the downward-pointing triangles of the Hamiltonian
themselves form a triangular lattice, and for each site on this lattice
which corresponds to a trio of spins of which either one or three are up,
there is an energy contribution of $J$ to the Hamiltonian.  Thus we can
represent each state of the lattice by a set of defect variables
\begin{equation}
d_i = (s_i+s_j+s_k) \bmod 2,
\label{defects}
\end{equation}
which take the value 1 when a defect is present and 0 otherwise.  In terms
of these defect variables, the Hamiltonian takes the form of a set of
non-interacting Ising spins in an external field $J$:
\begin{equation}
H = J \sum_i d_i.
\label{defsh}
\end{equation}
This simple form for the Hamiltonian allows us to solve for the model's
equilibrium behavior exactly.  However, there is a price to be paid for
this simplicity in terms of an increased complexity in the dynamics.  In
the defect representation of the model, a single spin-flip corresponds to
flipping the states of three defects at the vertices of an {\em
  upward}-pointing triangle.  Thus our model displays clearly the duality
between dynamics and interactions which is present in all systems; we can
think of it either as a system of interacting spins with single-spin-flip
dynamics, or as non-interacting spins with a constrained dynamics in which
we flip three spins at once.  In the remainder of the paper, we will for
the most part adopt the latter description.

The first step in solving for the equilibrium partition function of the
model is to find the set of allowed configurations of the defect variables
$d_i$, so that we can perform the sum over them.  Clearly the number of
sites on the defect lattice is the same as that on the spin lattice and
hence the maximum possible number of defect configurations is the same as
the number of spin configurations of our original spin variables.  We now
show that, for certain boundary conditions, there is a one-to-one
correspondence between spin configurations and defect configurations.

Consider three spin configurations $\lbrace s^{(0)} \rbrace$, $\lbrace
s^{(1)} \rbrace$ and $\lbrace s^{(2)} \rbrace$, related as follows:
\begin{equation}
s^{(2)}_i = (s^{(1)}_i + s^{(0)}_i) \bmod 2.
\label{spinsum}
\end{equation}
The corresponding defect configurations are similarly related:
\begin{equation}
d^{(2)}_i = (d^{(1)}_i + d^{(0)}_i) \bmod 2.
\label{defectsum}
\end{equation}
If spin configurations~1 and~2 are to have the same defect configuration
$d^{(1)}_i = d^{(2)}_i$ for all $i$, it follows that the defect variables
corresponding to configuration~0 must all be zero, i.e.,~that
configuration~0 must be a ground state of the system.  If we can show that
there is only one such ground state---the trivial one in which all spins
are zero---then it follows that $\lbrace s^{(1)} \rbrace$ and $\lbrace
s^{(2)} \rbrace$ are identical and the mapping of spin states to defect
states is one-to-one.  We can indeed show this in the case of a lattice
which has length $L=2^k$ for integer $k$ along one dimension and periodic
boundary conditions.  The argument runs as follows.

Suppose we have a lattice in the form of a rhombic strip of width $L=2^k$.
If the configuration is to be a ground state, then there can be no defects
at any site on the lattice.  This allows us to calculate the values of the
spins in one row given those in the preceding row since, by
Equation~\eref{defects}, each one must be the sum mod~2 of the two above
it.  If $s_{ij}$ is the $j^{\rm th}$ spin of the $i^{\rm th}$ row, then
\begin{equation}
s_{i+1,j} = (s_{ij} + s_{i,j+1}) \bmod 2,
\label{ca}
\end{equation}
where $s_{i+1,j}$ is the site below $s_{ij}$ and $s_{i,j+1}$.
The spins in the next row after this are then
\begin{eqnarray}
s_{i+2,j} &=& (s_{ij} + 2 s_{i,j+1} + s_{i,j+2}) \bmod 2\nonumber\\
          &=& (s_{ij} + s_{i,j+2}) \bmod 2.
\end{eqnarray}
By iterating this argument it can now be shown that a similar result
applies for each row which is a power of 2 away from the initial one.  For
$L$ a power of~2, we then have
\begin{equation}
s_{i+L,j} = (s_{ij} + s_{i,j+L}) \bmod 2 
          = (2s_{ij}) \bmod 2 = 0,
\end{equation}
since $s_{i,j+L} = s_{ij}$ because of the boundary conditions.  Given that
both $i$ and $j$ are arbitrary, it immediately follows that every spin on
the lattice is zero.

\begin{figure}[t]
\begin{center}
\psfig{figure=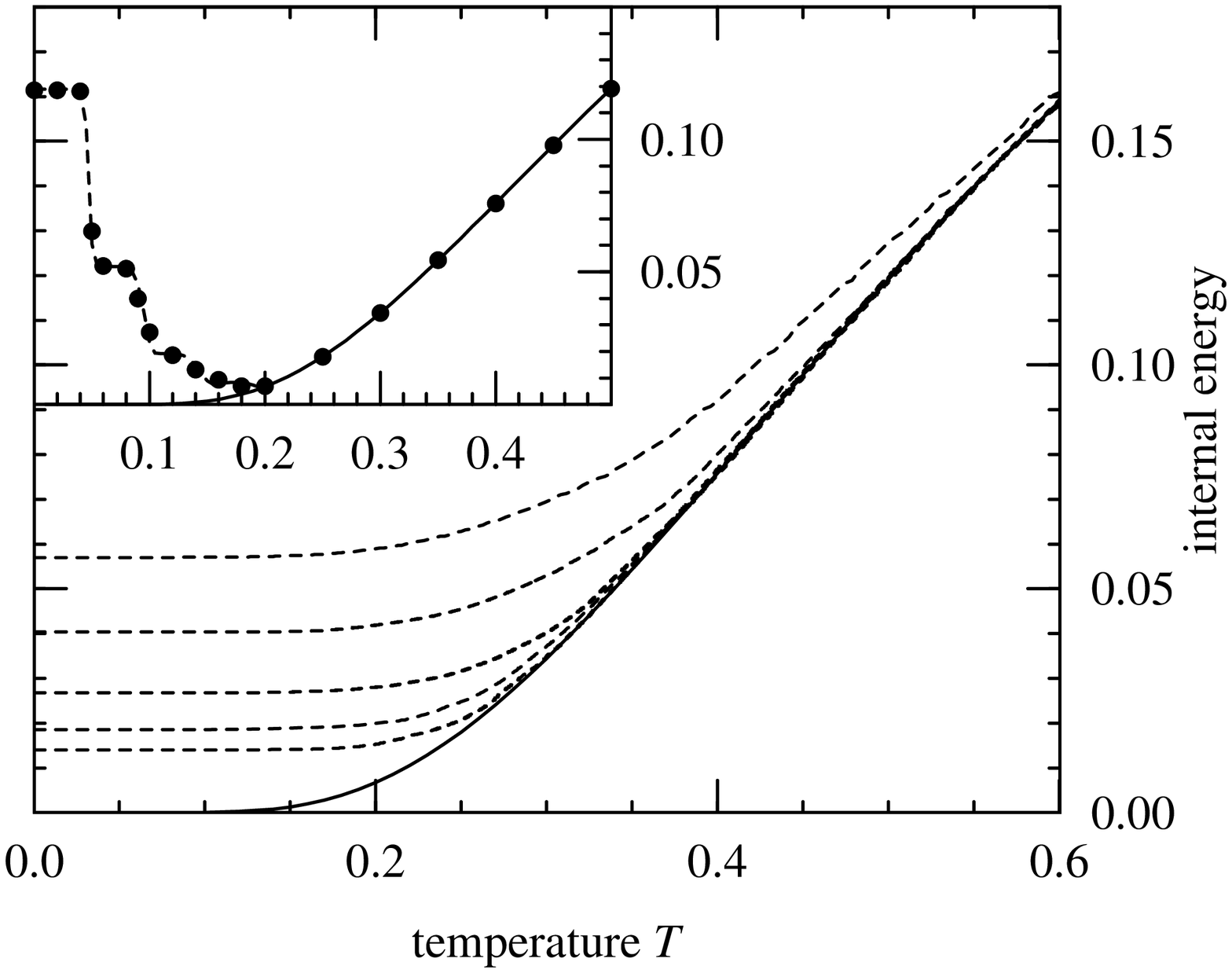,width=4.5in}
\end{center}
\capt{The internal energy per site as a function of temperature in
  a series of annealing simulations using an exponential cooling schedule
  (dashed lines) compared to the exact solution at equilibrium (solid
  line).  Inset: the internal energy following a quench from $T=\infty$ to
  a finite temperature.  The points are data from Monte Carlo simulations,
  the solid line is the equilibrium solution, and the dashed line is a fit
  of the form~\eref{fit}.  The steps in the fitted function correspond to
  the time-scales $\tau_k$, Equation~\eref{tauk}.  Their heights are set by
  the fit parameters $A_k$, but their positions are absolute.}
\label{et}
\end{figure}

Thus we have demonstrated that, for lattices of length a power of two along
at least one dimension, there is a unique ground state in which all spins
have value zero, which in turn implies a one-to-one mapping of defect
states to spin states.  Given the Hamiltonian~\eref{defsh}, the partition
function of the model is then simply
\begin{equation}
Z = \sum_{n=0}^N {N \choose n} \,{\rm e}^{-\beta J n}
  = [1 + {\rm e}^{-\beta J}]^N.
\label{partition}
\end{equation}
The equilibrium internal energy per site is then
\begin{equation}
E_{\rm eq} = -{1\over Z} {\partial Z\over\partial\beta}
           = {J\over 1+{\rm e}^{\beta J}}.
\label{equil}
\end{equation}
For lattice sizes which are not a power of 2, the proof above no longer
applies and more than one ground state may exist\cite{Note1}.  In that
case, not all defect configurations can occur.  However, the ones that do
exist all correspond to the same multiplicity of spin configurations, one
for each ground state.  Since the states of the spins in a particular
ground state are determined by the spins on any one row of the lattice, the
number of ground states can increase at most as $N_\ground \sim {\rm e}^L
\simeq {\rm e}^{\sqrt{N}}$ with lattice size.  In addition, all of the
defect states can be chosen independently except for those on one row,
which may be restricted to some extent by the requirement that the spin
configuration to be consistent with the periodic boundary conditions.  This
means that the partition function can be written as a sum
\begin{equation}
Z = N_\ground \,\sum_{n=0}^{N-L} {N-L \choose n} 
              \,{\rm e}^{-\beta J (n+\delta n)}
\end{equation}
where $\delta n$ is the number of additional defects in that row determined
by our $N-L$ choices in the other rows.  Since $\delta n \le L \simeq
\sqrt{N}$, logarithmic derivatives of $Z$, and therefore bulk properties of
the system, converge to those of Equation~\eref{partition} for large $N$.

In Figure~\fref{et} we show our solution for the internal energy as a
function of temperature (solid line), along with Monte Carlo results from
the simulation of the model (dashed lines).  The simulations were performed
on a $128\times128$ rhombic system with $J=1$ using a Bortz-Kalos-Lebowitz
continuous time algorithm\cite{BKL75}.  Each curve represents the internal
energy as a function of temperature during an annealing experiment using an
exponential cooling schedule $T=T_0 {\rm e}^{-\gamma t}$ with $T_0=1$ and
cooling rates (top to bottom) of $\gamma=10^{-2}$, $10^{-3}$, $10^{-4}$,
$10^{-5}$, and $10^{-6}$ in units of inverse Monte Carlo steps per spin.
As the figure shows, the model's behavior is in good agreement with the
equilibrium solution at high temperatures, but falls out of equilibrium at
lower and lower temperatures as the cooling rate is decreased, in a manner
characteristic of glassy systems.

In the inset, we show the results of numerical experiments in which
the same system is quenched from $T=\infty$ to a fixed finite
temperature.  Each point represents the final average internal energy
of the system after $10^9$ Metropolis Monte Carlo steps per site
(i.e., more than $10^{13}$ steps total).  As we can see, the exact
solution is again in good agreement with the simulations for high
temperatures, but fails badly as $T \to 0$.

\section{The origin of glassiness in the model}
\label{glassy}
We can gain some insight into the model's loss of ergodicity if we
recall that the flipping of a single spin corresponds to flipping the
states of three defects in an upward-pointing triangle.  In the limit
of low temperature only those moves which flip the defects in
triangles containing either two or three defects are energetically
possible.  Triangles with one defect only will be exponentially
unlikely to change, and become local minima at $T=0$.  Hence there
will be a finite energy, and entropy, at $T=0$ \cite{Note2}.

In order to demonstrate that our model is truly glassy in the conventional
sense, however, we need to treat the finite temperature case and
investigate the distribution of energy barriers.  As we have demonstrated
above, the model has only one ground state, in which there are no defects
and all spins are zero.  We now show that the elementary excitations of the
model---those states lying closest to the ground state---are trios of
defects at the vertices of an upward-pointing equilateral triangle of
length $\ell=2^k$ on a side with integer $k$.

Equation~\eref{ca} tells us that the spins below an isolated defect form a
Pascal's triangle mod~2.  If we take a finite region of the lattice in the
form of an upward-pointing equilateral triangle, each defect in it produces
such a Pascal's triangle.  Then if the spins along the top sides of the
triangle are zero, the bottom row is the sum mod~2 of the corresponding
rows of each of the triangles.  We call this row of spins the {\em shadow}
of the region's defects.  The sum of the Pascal's triangles of an
upward-pointing triangle of three adjacent defects is zero; thus an move
that flips all three conserves the shadow, and one defect configuration can
be reached from another by a series of local moves if and only if they have
the same shadow.  In particular, only configurations with a zero shadow can
be local excitations of the ground state.  It is then straightforward to
show that no configurations with one or two defects can have a zero shadow,
and that the only such configurations with three defects are those arranged
in an upward-pointing triangle of side $2^k$.

\begin{figure}[t]
\begin{center}
\psfig{figure=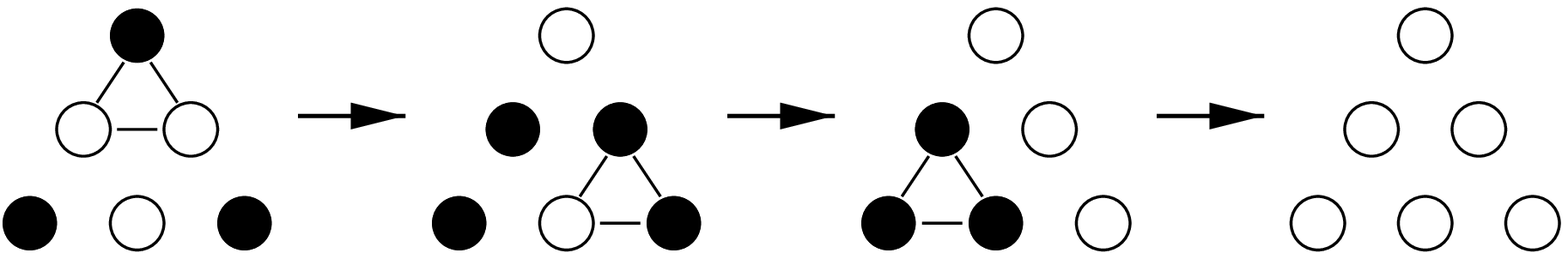,width=4.5in}
\end{center}
\capt{A triangle of side~$2^k$ can be flipped by flipping three triangles 
  of side~$2^{k-1}$.  The solid circles represent the defects and the lines
  indicate the triangles to be flipped at each step.}
\label{moves}
\end{figure}

Next, we ask what the energy barrier is for flipping a triangular
excitation of a given size.  The minimum-energy path for flipping a
triangle of side $2^k$ involves flipping three triangles of side $2^{k-1}$
in series, as shown in Figure~\fref{moves}.  Since the intermediate state
on this path has four defects rather than three, the total energy barrier
for the process is $J$ higher than that for flipping a triangle of half the
size.  This in turn is $J$ higher than the barrier for flipping triangles
of half {\em that\/} size, and so on, down to triangles of side~1 which
have barrier zero.  Thus the total height of the barrier which must be
crossed in order to create or remove a triangular excitation of side
$\ell=2^k$ is $J \log_2 \ell = kJ$, increasing logarithmically with
size\cite{Note3}.

In a system of linear dimension $L<2^{k+1}$, the largest possible
excitation is a triangle of side $2^k$, and $kJ \simeq J \log_2 L$ is the
largest energy barrier the system must cross to achieve ergodicity.  The
conventional view is that a glassy system should have energy barriers which
scale as a power of $L$.  Since $\log L$ is a limiting case of the power
law when the exponent tends to zero, our model can be considered marginally
glassy.  At low temperatures, assuming an Arrhenius law $\tau \propto
{\rm e}^{\beta\,\Delta E}$, the correlation time goes as
\begin{equation}
\tau \sim {\rm e}^{\beta J\log_2 L} = L^{\beta J/\ln 2}.
\label{corrtime}
\end{equation}
i.e.,~as a power-law in the system size, with the exponent increasing
linearly with $\beta$.  At high temperature, the fact that there are
several pathways for annealing away a triangular excitation reduces the
free energy barrier somewhat, but we believe that there is no sharp glass
transition.

\begin{figure}[t]
\begin{center}
\psfig{figure=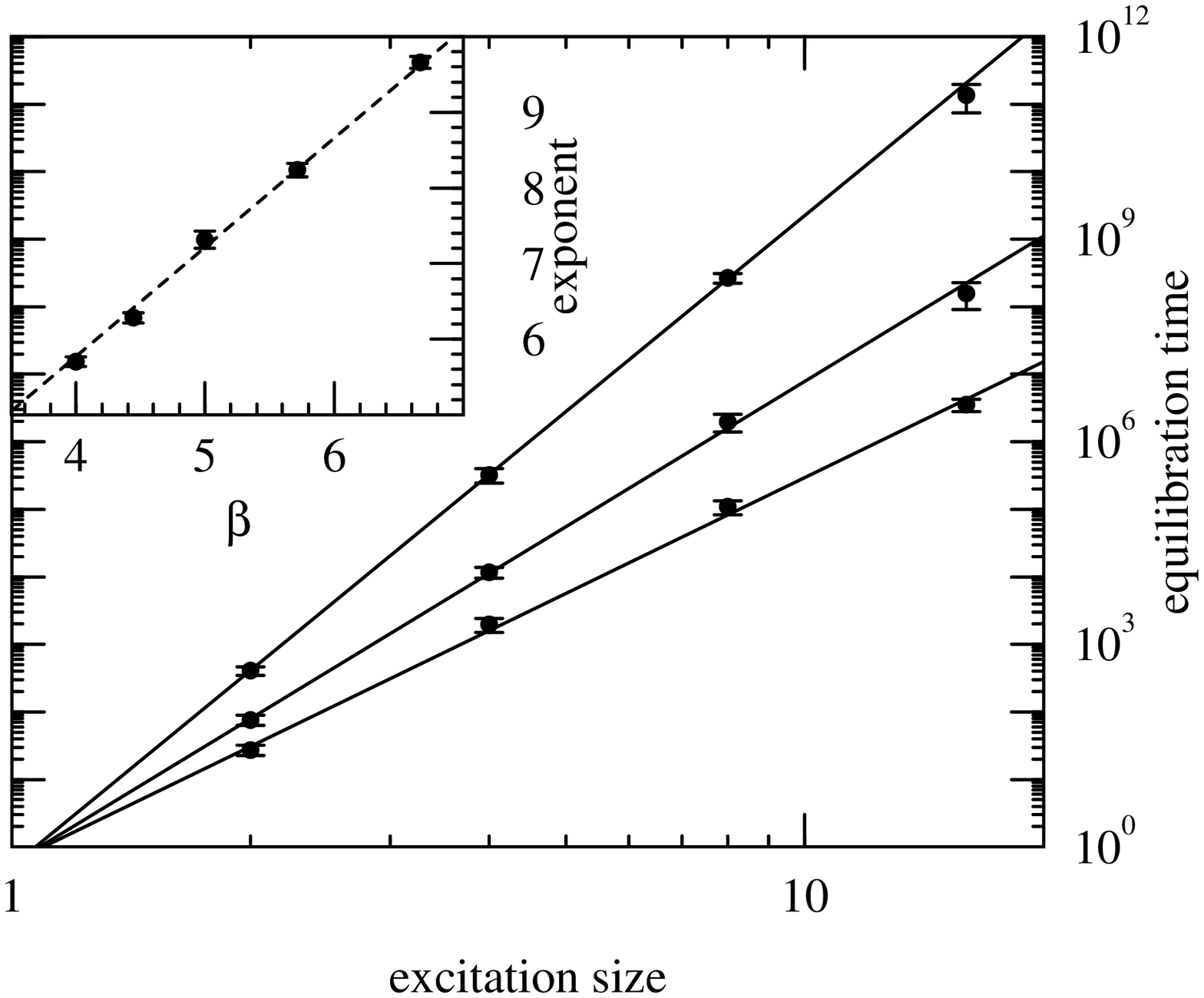,width=4.5in}
\end{center}
\capt{The time it takes the system to eliminate a single triangular 
  excitation of size $\ell=2$, $4$, $8$, $16$ for temperatures $T=0.15$,
  $0.20$, $0.25$.  Each set of points follows the expected power-law.
  Inset: the exponent of the power-law as a function of inverse temperature
  $\beta$.  The predicted value of $\beta J/\ln 2$,
  Equation~\eref{corrtime}, is shown as the dotted line.}
\label{st}
\end{figure}

We have confirmed these results in simulations of our model.  In
Figure~\fref{st} we show the time taken to equilibrate the system starting
from a state consisting of a single triangular excitation of a given size
for three different temperatures with $J=1$.  The expected power-law is
obeyed closely.  The lines should cross at the origin, since the time to
get rid of an excitation of size $\ell=1$ is unity, and to a reasonable
approximation they do this.  The exponent of the power law is shown as a
function of $\beta$ for five different temperatures in the inset figure.
The expected value of $\beta J/\ln 2$ is shown as the dotted line and
agrees well with our measurements.

We are now also in a position to explain the form taken by the Monte Carlo
results in Figure~\fref{et}.  Writing the time-scale for equilibration on
length-scales up to $\ell=2^k$ as
\begin{equation}
\tau_k = {\rm e}^{\beta Jk},
\label{tauk}
\end{equation}
we can write the energy of the system after time $t$ as
\begin{equation}
E(t) = E_{\rm eq} + \sum_k A_k {\rm e}^{-t/\tau_k},
\label{fit}
\end{equation}
where the quantities $A_k$ are temperature-independent constants.  The
dashed line in the inset of Figure~\fref{et} is of this form with $E_{\rm
  eq}$ taken from Equation~\eref{equil}, $t=10^9$ as in the simulations,
and the $A_k$ assigned by a least squares fit to the data.  Of particular
interest are the `steps' visible in this fit.  The temperatures $T_k$ at
which these occur are solutions of $\tau_k=t$:
\begin{equation}
T_k = Jk/\ln t,
\label{steps}
\end{equation}
with $k$ taking integer values up to $\log_2 L$.  Thus the temperature at
which the system fails to equilibrate is inversely proportional to the
logarithm of the cooling time $t$.  This goes to zero more slowly than any
power law as $t$ goes to infinity.

\section{Aging}
\label{aging}
\begin{figure}[t]
\begin{center}
\psfig{figure=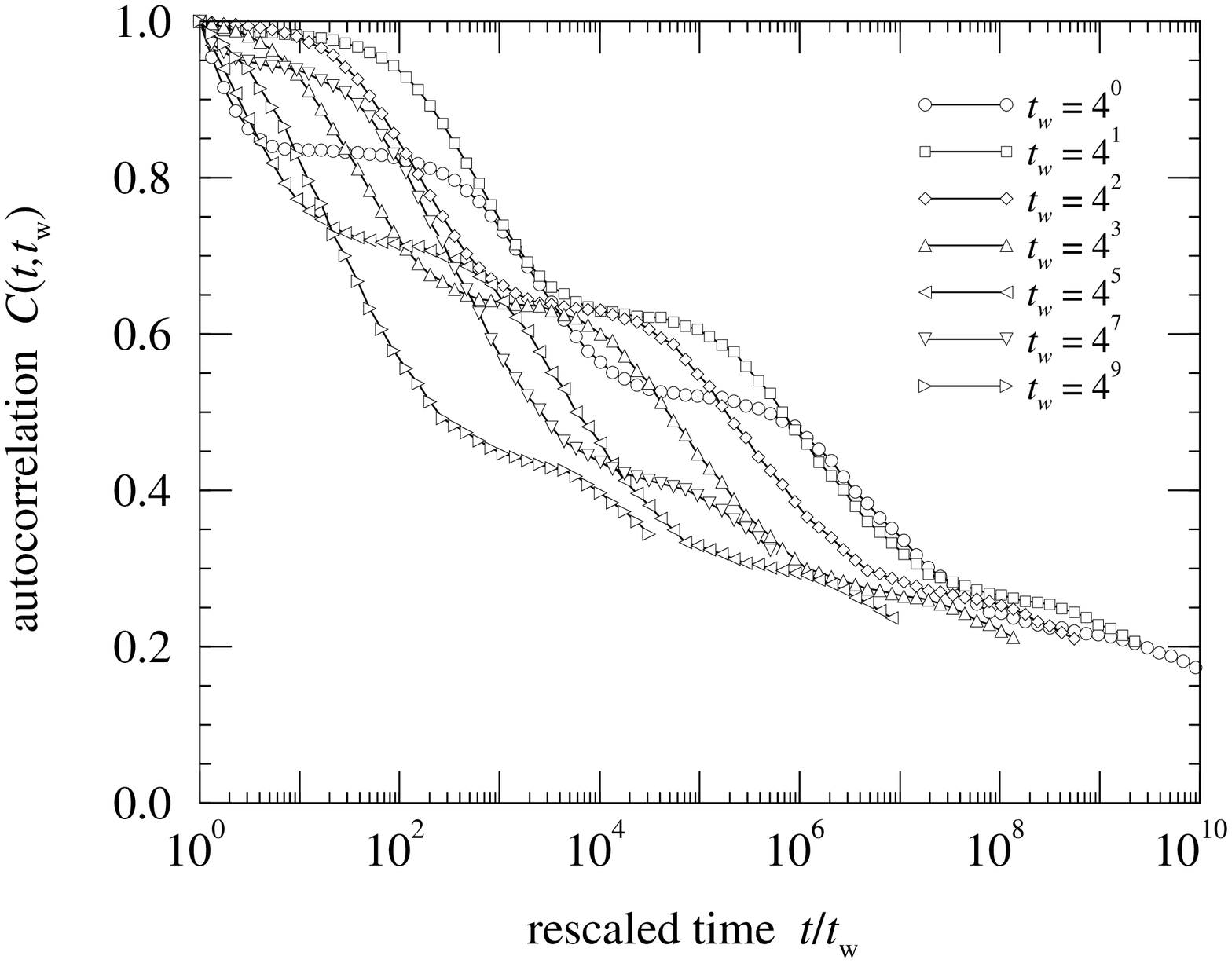,width=4.5in}
\end{center}
\capt{The two-time correlation function $C(t_w,t)$ plotted as a function
  of $t/t_w$ for $t_w=4^n$ with $n=0,1,2,3,5,7,9$.  Note that the horizontal
  (time) axis is logarithmic.}
\label{correl}
\end{figure}

We have also looked at the aging behavior of the model by examining the
behavior of the two-time spin-spin connected correlation function
$C(t_w,t)$ as a function of waiting time $t_w$.  This function is defined
by
\begin{equation}
C(t_w,t) = \overline{s_i(t_w) s_i(t)} -
\overline{s_i(t_w)}\>\>\overline{s_i(t)},
\end{equation}
where the bar indicates an average over the lattice.  If a system relaxes
to equilibrium exponentially fast, $C$ is a function only of $t-t_w$.  In
our model however, as is typical in systems with slow relaxation, $C$
depends on $t_w$.  In Figure~\fref{correl} we show $C$ as a function of the
ratio $t/t_w$ for a variety of values of $t_w$.  The figure has a number of
notable features.  The `steps' in the correlation function arise because
all barriers in the model are multiples of $J$.  This is true in some other
glassy models as well, such as the Edwards--Anderson Ising
spin-glass\cite{FH91,MPR97} with random bonds $\pm J$.  However, in that
model, the height of the highest barrier, and hence the density of steps
per unit volume, increases as a power of the size of the system, so that
for a system of moderate size, the steps in $C(t_w,t)$ are small enough to
be indistinguishable to the eye.  In our model the height of the highest
barrier in the system increases only logarithmically with system size, so
that the steps are still visible even for quite large lattices.

Ignoring the steps in the correlation function, the figure shows that the
rate of decline of the correlation function as a function of $t/t_w$ is
roughly independent of $t_w$, although there is no actual collapse of the
curves onto one another as there is in some other models\cite{KRS94}.  This
is precisely the type of behavior which one would expect to see in this
system, since the energy barriers we need to cross at each succeeding
length scale $\ell = 2^k$ are a constant amount $J$ higher than those at
the previous one, so that the corresponding time-scales increase by a
constant factor (see Equation~\eref{tauk}).  (By contrast, a plot in which
the time is not scaled by the factor $t_w$ gives no collapse of the
correlation function, whereas in a non-glassy system such a plot should
collapse perfectly.)  Our model is instructive in this respect, since it
makes the origins of the aging behavior particularly clear.

\section{Conclusions}
\label{concs}
To conclude, we have introduced a two-dimensional spin model with no
randomness and only short-range interactions.  Under single-spin-flip
dynamics it displays glassy behavior, with barrier heights growing
logarithmically with system size.  We have given an exact solution for
both the equilibrium properties of the model and the distribution of
energy barriers.  We have performed numerical simulations which
confirm our analytic results to within the available precision.  The
model seems to have no sharp glass transition, and falls out of
equilibrium at a temperature which decreases logarithmically as a
function of the cooling time.  It also displays clear aging behavior
consistent with our understand of the distribution of energy barriers.

\section*{Acknowledgements}
The authors would like to thank Nelson Minar for performing some of the
early simulations of this system, and Richard Palmer, Heiko Rieger, David
Sherrington, and Paolo Sibani for useful discussions.  This work was funded
in part by the Santa Fe Institute and DARPA under grant number ONR
N00014--95--1--0975.


\begin{references}
%
\bibitem{Jackle86}
{\frenchspacing J. J\"ackle, Rep. Prog. Phys. {\bf 49}, 171 (1986).}
%
\bibitem{FH91}
{\frenchspacing K. H. Fischer and J. A. Hertz, {\em Spin Glasses,}
  Cambridge University Press, Cambridge (1991).}
%
\bibitem{Parisi94}
{\frenchspacing G. Parisi, {\it Field Theory, Disorder and Simulations,}
  World Scientific, Singapore (1994).}
%
\bibitem{MPR97}
{\frenchspacing E. Marinari, G. Parisi, and J. J. Ruiz-Lorenzo, in {\it
Spin Glasses and Random Fields,} P. Young (ed.), World Scientific,}
Singapore (1997).
%
\bibitem{NB99}
{\frenchspacing M. E. J. Newman and G. T. Barkema, {\em Monte Carlo Methods
    in Statistical Physics,} Oxford University Press (1999).}
%
\bibitem{SK75}
{\frenchspacing D. Sherrington and S. Kirkpartrick,
  Phys. Rev. Lett. {\bf35}, 1972 (1975).}
%
\bibitem{Derrida81}
{\frenchspacing B. Derrida, Phys. Rev. B {\bf24}, 2613 (1981).}
%
\bibitem{Ritort95}
{\frenchspacing F. Ritort, Phys. Rev. Lett. {\bf75}, 1190 (1995).}
%
\bibitem{SHS92}
{\frenchspacing J. D. Shore, M. Holzer, and J. P. Sethna, Phys. Rev. B
  {\bf 46}, 11376 (1992).}
%
\bibitem{Rieger92}
{\frenchspacing H. Rieger, Physica A {\bf184}, 279 (1992).}
%
\bibitem{KRS94}
{\frenchspacing J. Kisker, H. Rieger, and H. Schreckenberg, J. Phys. A {\bf27},
  L853 (1994).}
%
\bibitem{bernasconi}
{\frenchspacing J. Bernasconi, J. Physique {\bf 48}, 559 (1987).}
%
\bibitem{bouchaud}
{\frenchspacing J.P. Bouchaud and M. M\'ezard, J. Phys. I France {\bf 4},
1109 (1994).}
%
\bibitem{marinari}
{\frenchspacing E. Marinari, G. Parisi and F. Ritort, J. Phys. A
{\bf 27}, 7647 (1994).}
%
\bibitem{BW73}
{\frenchspacing R. J. Baxter and F. Y. Wu, Phys. Rev. Lett. {\bf31}, 1294
  (1973).}
%
\bibitem{BNB94}
  {\frenchspacing G. T. Barkema, M. E. J. Newman, and M. Breeman, Phys.
  Rev. B {\bf50}, 7946 (1994).}
%
\bibitem{MRRTT53}
{\frenchspacing N. Metropolis, A. W. Rosenbluth, M. N. Rosenbluth,
  A. H. Teller, and E. Teller, {\em J. Chem. Phys.} {\bf 21}, 1087 (1953).}
%
\bibitem{Note1}
  Since Equation~\eref{ca} is the rule for iterating a certain additive
  cellular automaton in one dimension, we can count ground states with
  periodic boundary conditions by counting periodic orbits of this cellular
  automaton.  The number of orbits as a function of lattice size turns out
  to have interesting number-theoretic properties, and has been studied by
  O.~Martin, A.~M.~Odlyzko, and S.~Wolfram, {\em Comm.\ Math.\ Phys.} {\bf
  93}, 219 (1984).
%
\bibitem{BKL75}
{\frenchspacing A. B. Bortz, M. H. Kalos, and J. L. Lebowitz, 
  {\em J. Comp. Phys.} {\bf 17}, 10 (1975).}
%
\bibitem{Note2} The set of local minima in our model is the same as
  the set of allowed states of Baxter's hard-hexagon model
  (R.~J.~Baxter, {\em Exactly Solved Models in Statistical Mechanics,}
  Academic Press, London, 1982).  However, the cooling process creates
  spatial correlations, giving an entropy at $T=0$ lower than that of
  Baxter's model.
%
\bibitem{Note3}
  We can generalize this to higher dimensions.  For example, in $d=3$ 
  the excitations of a model with four-spin interactions on tetrahedra 
  of one orientation on a fcc lattice will be tetrahedra of size $2^k$ with 
  energy barriers $2kJ$.  A model with interactions on both kinds of
  tetrahedra is discussed in \cite{Rieger92}.
%
\end{references}
\end{document}